\documentclass[conference]{IEEEtran}
\IEEEoverridecommandlockouts

\usepackage{amsmath,graphicx}

\makeatletter
\let\NAT@parse\undefined
\makeatother

\usepackage{hyperref}
\usepackage{lipsum}
\usepackage{booktabs}
\usepackage{float}
\usepackage{mdframed}
\usepackage{makecell}

\usepackage{cite}
\usepackage{amsmath,amssymb,amsfonts}
\usepackage{algorithmic}
\usepackage{graphicx}
\usepackage{textcomp}
\usepackage{xcolor}
\def\BibTeX{{\rm B\kern-.05em{\sc i\kern-.025em b}\kern-.08em
    T\kern-.1667em\lower.7ex\hbox{E}\kern-.125emX}}

\title{Improving Automatic Speech Recognition for Speakers Treated for Oral Cancer using Data Augmentation and LLM Error Correction}

\makeatletter
\newcommand{\linebreakand}{%
  \end{@IEEEauthorhalign}
  \hfill\mbox{}\par
  \mbox{}\hfill\begin{@IEEEauthorhalign}
}
\makeatother

\author{\IEEEauthorblockN{Hidde Folkertsma}
\IEEEauthorblockA{\textit{University of Groningen} \\
Groningen, The Netherlands \\
hiddefolkertsma@gmail.com \\ 
\\                           
}
\and
\IEEEauthorblockN{Thomas B. Tienkamp}
\IEEEauthorblockA{\textit{University Hospital Cologne} \\
Cologne, Germany \\
thomas.tienkamp@uk-koeln.de}
\and
\IEEEauthorblockN{Sebastiaan A.H.J. de Visscher}
\IEEEauthorblockA{\textit{University Medical Center Groningen} \\
\textit{University of Groningen} \\
Groningen, The Netherlands \\
s.a.h.j.de.visscher@umcg.nl
}
\linebreakand
\IEEEauthorblockN{Max J.H. Witjes}
\IEEEauthorblockA{\textit{University Medical Center Groningen} \\
\textit{University of Groningen} \\
Groningen, The Netherlands \\
m.j.h.witjes@umcg.nl}
\and
\IEEEauthorblockN{Rob J.J.H. van Son}
\IEEEauthorblockA{\textit{Netherlands Cancer Institute} \\
\textit{University of Amsterdam} \\
Amsterdam, The Netherlands \\
r.v.son@nki.nl}
\and
\IEEEauthorblockN{Jiapan Guo}
\IEEEauthorblockA{\textit{University of Groningen} \\
Groningen, The Netherlands \\
j.guo@rug.nl}
\linebreakand
\IEEEauthorblockN{Bence Mark Halpern}
\IEEEauthorblockA{\textit{Nagoya University} \\
Nagoya, Japan \\
halpern.bence.e8@f.mail.nagoya-u.ac.jp}}

\begin{document}
\maketitle
\thispagestyle{empty}
\pagestyle{empty}

\begin{abstract} 
In recent years, the performance of automatic speech recognition (ASR) systems has made considerable progress. Unfortunately, for people with speech impairments, such as people treated for oral cancer (OC), ASR performance is still lagging behind. The scarcity and variability of OC speech data makes development of ASR models for this type of speech difficult. In this work, we use data augmentation and large language model (LLM) error correction to mitigate this problem. 
We apply various augmentation techniques on a corpus of Dutch oral cancer speech to create synthetic data, and evaluate their effect on ASR performance. We finetune Whisper and Massively Multilingual Speech (MMS) models for each augmentation technique and observe, on average, an 8\% relative decrease in Word Error Rate (WER) when including data created using text-to-speech (TTS).
When employing LLMs for error correction, we see a further 21.4-26.2\% relative decrease in WER for finetuned ASR models and a 10.0\% relative decrease for non-finetuned models. Overall, we achieve a 40\% relative WER decrease for Whisper and a 50\% relative WER decrease for MMS, indicating that a combination of data augmentation and LLM correction is a viable strategy for the recognition of OC speech.
\end{abstract}


\section{Introduction} \label{sec:introduction} 
Oral cancer is a disease that affects around 500,000 people per year worldwide \cite{shieldGlobalIncidenceLip2017, ferlayCancerStatisticsYear2021}. Most oral cancers occur on the tongue or the floor of the mouth, and are surgically resected (i.e., removed) during treatment \cite{constantinescuSpeechDeficitsAssociated2019}. Resecting the tumor often involves partial resection and reconstruction of the tongue, which can lead to speech problems. Unless the patient also undergoes radiation treatment, other anatomical structures involved in speech production (e.g., the larynx) remain untouched. People treated for oral cancer therefore primarily experience articulatory problems as opposed to voice problems. The degree of speech impairment varies based on tumor size and site, tongue reconstruction technique and post-operative tongue mobility \cite{bressmannSpeechDisordersRelated2021}. To maintain speech intelligibility, people treated for oral cancer may employ varying compensatory strategies \cite{tienkampArticulatoryKinematicChanges2025}.

Owing to the above factors, post-operative speech outcomes can vary significantly between speakers, which makes automatic speech recognition (ASR) of oral cancer speech a challenging task. ASR systems are usually trained on large amounts of typical speech \cite{latifTransformersSpeechProcessing2025}, and have been shown to generalize poorly to pathological speech, such as that of people treated for oral cancer \cite{halpernLowresourceAutomaticSpeech2022, wildenburgAutomaticSpeechRecognition2022} or people with dysarthria \cite{sheikhDeepLearningPathological2025, qianSurveyTechnologiesAutomatic2023}. Unfortunately, training ASR systems from scratch on oral cancer speech data is currently not feasible due to the scarcity of available data. Pathological speech data is difficult to collect for several reasons, such as the relatively few available speakers, and speakers having issues speaking for prolonged periods of time.

To overcome data scarcity, adding synthetic data to the training set, also known as data augmentation, has been proposed as a fruitful strategy. To date,  traditional signal processing methods such as time stretching, speed perturbation \cite{koAudioAugmentationSpeech2015} and vocal tract length perturbation \cite{jaitlyVocalTractLength2013} have been used to adjust typical speech to more closely match the spectrotemporal characteristics of pathological speech. More recent research has also investigated text-to-speech (TTS) and voice conversion (VC) systems for data augmentation. Compared to the signal processing methods, TTS and VC aim to more accurately reconstruct the spectrotemporal characteristics (e.g., articulatory imprecision) of the pathological speech, thus creating more realistic-sounding synthetic pathological speech.

Data augmentation for pathological speech recognition has been studied widely for dysarthria, using traditional methods \cite{vachhaniDataAugmentationUsing2018, gengInvestigationDataAugmentation2020, gracelliExploringAlternativeData2024, naeiniImprovingDysarthricSpeech2024}, VC-based methods \cite{elhajalUnsupervisedRhythmVoice2025, liInclusiveASRInvestigating2025} and TTS-based methods \cite{soleymanpourSynthesizingDysarthricSpeech2022, hermannFewshotDysarthricSpeech2023, leungTrainingDataAugmentation2024}. Results indicate that all methods can improve ASR performance. However, training VC and TTS models faces the same data scarcity issue as ASR models, leading to a chicken-and-egg problem. Zero-shot and few-shot VC and TTS models, while useful for generating augmented samples for arbitrary speakers, require large amounts of training data, which is not feasible for pathological speech. Other VC/TTS paradigms, for example with a fixed target speaker, are limited in the extra variety they add to the augmented dataset.

To the best of our knowledge, there is no prior research on data augmentation for oral cancer speech recognition. However, both ASR \cite{halpernLowresourceAutomaticSpeech2022, wildenburgAutomaticSpeechRecognition2022} and VC \cite{halpernImprovingSeverityPreservation2023} have been studied. Halpern et al. \cite{halpernLowresourceAutomaticSpeech2022} trained baseline models on the WSJ corpus \cite{paul-baker-1992-design}, and then finetuned these on a dataset of spontaneous English oral cancer speech from YouTube. The finetuned models showed a 10\% relative decrease in Word Error Rate (WER) when evaluated on oral cancer speech. Wildenburg \cite{wildenburgAutomaticSpeechRecognition2022} trained an ASR model on Dutch typical speech from the Corpus Gesproken Nederlands \cite{oostdijkHetCorpusGesproken1999}, and evaluated on NKI-RUG-UMCG \cite{halpernManipulationOralCancer2022}, a dataset of Dutch oral cancer speech. Their model achieves a WER of 17.4\% for the typical speech in NKI-RUG-UMCG and a WER of 62.3\% for the oral cancer speech.

Another promising approach for improving ASR performance is asking a large language model (LLM) to correct the ASR model's predictions. ASR models are prone to directly transcribing articulatory deviations of speakers, and we expect that the strong linguistic knowledge of LLMs can correct these mistakes, improving recognition performance. This approach has recently been found beneficial for dysarthric speech \cite{heRobustDysarthricSpeech2025}. However, LLM error correction has remained unexplored for oral cancer speech recognition.

In this work, we explore using both traditional signal processing methods as well as VC and TTS methods to perform data augmentation on NKI-RUG-UMCG. We use the augmented variants of the dataset to finetune Whisper \cite{radfordRobustSpeechRecognition2022} and Massively Multilingual Speech (MMS) \cite{pratapScalingSpeechTechnology2024} models, and provide a comparative analysis of how each augmentation method affects the ASR models' performance. To further understand how the augmented data affects ASR performance, we finetune the MMS models both with and without a language model, and finetune the Whisper models both fully as well as using parameter-efficient finetuning (LoRA) \cite{huLoRALowRankAdaptation2022}. Finally, we investigate how using large language models as an error-correcting post-processing step can further affect performance.

\section{Augmentation methods}

\subsection{Time stretching}
Time stretching (TS), also known as tempo stretching, is a method to change the speed of an audio signal while maintaining its pitch. Applying time stretching to typical speech can be used to convert it to an atypical speech rate, which is commonly seen in pathological speech. In the context of dysarthria, this has led to improved results \cite{vachhaniDataAugmentationUsing2018, bhatImprovedASRPerformance2022}, indicating it might be promising for oral cancer speech as well.

Time stretching can be implemented by performing a short-time Fourier transform on the signal, and then inverting the result with a smaller or greater distance between frames to respectively slow speed the signal up or slow it down. To prevent issues such as phase cancellation, methods such as phase vocoding are commonly employed as an intermediate step in the frequency domain. 

\subsection{Vocal tract length perturbation}
Vocal tract length, defined by the length between the vocal folds and the lips, has an effect on the vocal tract's resonance frequencies (e.g., vowel  formants). Shorter vocal tracts tend to produce higher formant frequencies, and longer vocal tracts produce lower ones. Vocal tract length perturbation (VTLP) \cite{jaitlyVocalTractLength2013} aims to mimic the effect of a change in vocal tract length, by linearly warping the frequency axis of a spectrogram. This has been shown to consistently improve performance for ASR on typical speech. As VTLP changes the resonance frequencies, it can be seen as a method to increase speaker variety. Since our dataset consists of a low number of speakers, adding speaker variety may help the model generalize to unseen speakers. Moreover, oral cancer surgery can change the shape of the vocal tract \cite{constantinescuSpeechDeficitsAssociated2019}. This can lead to atypical resonances in the oral cavity. VTLP may be useful to model such possible speech outcomes.

\subsection{Speed perturbation}
Speed perturbation (SP) is similar to time stretching, but does not preserve pitch information. This can be achieved by simply multiplying the sample rate information of a clip by the stretch factor, and then resampling back to the original sample rate. Speed perturbation changes both temporal and spectral characteristics, and can be seen as a combination of time stretching and VTLP \cite{koAudioAugmentationSpeech2015}.

\subsection{Voice conversion (kNN-VC)}
Voice conversion (VC) is the task of converting the speaker identity of an utterance to that of a target speaker, without changing the linguistic contents. We employ k-nearest neighbour voice conversion (kNN-VC) \cite{baasVoiceConversionJust2023} for our VC experiments. kNN-VC is a VC method that extracts a sequence of WavLM \cite{chenWavLMLargeScaleSelfSupervised2022} features from both source and target speech, and then replaces each feature in the source sequence by the average of the $k$ most similar features from the target sequence. This is a simple method that relies on WavLM's ability to encode features such as pitch and speaker identity. However, since the features are swapped one-for-one, the converted speech is almost identical in terms of tempo and pacing. As a result, kNN-VC cannot be expected to provide a larger variety in speech rhythm.

\subsection{Text-to-speech (XTTSv2)}
For our TTS experiments, we use XTTSv2 \cite{casanovaXTTSMassivelyMultilingual2024}, a state-of-the-art multilingual zero-shot TTS model, trained on typical speech. Since our training dataset contains a restricted number of prompts, linguistic variety is limited. TTS can help mitigate this, as any text prompt can be used to generate audio, thus increasing linguistic variety. This is a key advantage over other augmentation methods.

\section{Experimental setup}

\subsection{Dataset}
We conduct our experiments on NKI-RUG-UMCG \cite{halpernManipulationOralCancer2022}, a dataset of Dutch read speech. NKI-RUG-UMCG consists of 202 prompts, read by 11 patients (6 male, 5 female; age range = 46-76 years, mean age = 62 years) treated for oral cancer, and 8 age-matched controls (5 male, 3 female; age range = 56-69 years, mean age = 63 years). 5 patients have undergone tongue surgery, and 6 have undergone jaw surgery. All speakers are from the northern Netherlands. Each prompt is either a sentence from a short linguistically diverse story or a news headline from the Wablieft corpus \cite{vandeghinsteWablieftEasytoReadNewspaper2019}. The dataset is sampled at 22,050 Hz, but was resampled to 16 kHz for all experiments.

In accordance with similar work on dysarthric speech \cite{yueExploringAppropriateAcoustic2020, pranantaEffectivenessTimeStretching2022, hermannFewshotDysarthricSpeech2023, schuUsingUASpeechTORGO2023, soleymanpourAccurateSynthesisDysarthric2024}, we employ a leave-one-speaker-out (LOSO) approach. For each held-out patient speaker in our dataset, we finetune a model on all the other speakers, using the held-out speaker as the test set. For NKI-RUG-UMCG, this means each experiment consists of 11 finetuning runs.

\begin{table}[!b]
    \caption{Training dataset size and composition for each experiment.}
    \label{tab:dataset_augmented_sizes}
    \centering
    \begin{tabular}{lcccc} \toprule
    & \multicolumn{2}{c}{\textbf{Number of utterances (duration)}}\\
    \cmidrule(lr){2-3}
    \cmidrule(lr){4-5}
    \textbf{Dataset} & \textbf{Original} & \textbf{Augmented}\\\midrule
    All speakers (no augm.)  & 2496 (4.45 h) & -           \\
    Patients only (no augm.) & 1560 (2.89 h) & -           \\
    TS, SP                   & 1560 (2.89 h) & 936  (1.74 h) \\
    VTLP, VC                 & 1560 (2.89 h) & 1560 (2.89 h) \\
    TTS                      & 1560 (2.89 h) & 2020 (2.39 h) \\\bottomrule
    \end{tabular}
\end{table}
\autoref{tab:dataset_augmented_sizes} shows the number of utterances and average duration of our training datasets. Note that the exact durations of the splits vary slightly based on the current test speaker. The test set consists of 46 utterances, with an average duration of 5.6 minutes.

\begin{table}
\centering
\caption{Dataset prompt splits.}
\label{tab:dataset_prompt_splits}
\small
\resizebox{\columnwidth}{!}{
\begin{tabular}{rrc}
\toprule
\textbf{Name} & \textbf{Sentences} & \textbf{Split} \\
\midrule
Papa en Marloes (LIT1)          \cite{weijerNasaliteit1991}     & 8  & Test  \\
Man uit Finland (LIT2)          \cite{martensOntwikkeling2010}  & 14 & Train \\
De Noordenwind en de Zon (LIT3) \cite{vanSonIFA2001}            & 8  & Test  \\
Els gaat naar de markt (LIT4)   \cite{weijerNasaliteit1991}     & 10 & Train \\
Meneer van Dam (LIT5)           \cite{weijerNasaliteit1991}     & 6  & Train \\
Jorinde en Joringel (LIT6)      \cite{vanSonIFA2001}            & 80 & Train \\
Wablieft (NEWS)                 \cite{vandeghinsteWablieftEasytoReadNewspaper2019} & 76 & Test (1-30), \\&&Train (31-76) \\
\midrule
Total                                                           & 202 & (156 train, 46 test) \\
\bottomrule
\end{tabular}
}
\end{table}
For datasets where speakers read identical prompts, extra care must be taken to avoid linguistic overlap between the test and training splits \cite{yueExploringAppropriateAcoustic2020}. We therefore select a portion of the news headlines and two stories as the test set, and use the remainder for training. The splits are shown in \autoref{tab:dataset_prompt_splits}.

\subsection{Augmentation methods}
We use the phase vocoding-based time stretching implementation from Librosa \cite{mcfeeLibrosaAudioMusic2015}, applying a stretch factor of 0.9 to all control speakers. To perform speed perturbation, we use the SpeedPerturbation implementation from \texttt{torchaudio} \cite{yangTorchAudioBuildingBlocks2022}, again applying a stretch factor of 0.9 to all control speakers. 

For the the VTLP experiments, we employ the implementation from \texttt{nlpaug} \cite{ma2019nlpaug}, and apply it to all patient speakers, using the default settings.

For kNN-VC, we use its authors' implementation\footnote{~\url{https://github.com/bshall/knn-vc}} with the VAD trigger level set to 2 and $k = 8$. We randomly pair source speakers to a target control speaker. These pairs are identical for all kNN-VC experiments. 

For XTTS, we use the official implementation\footnote{~\url{https://github.com/coqui-ai/TTS}} with the language set to Dutch. We use the first 2222 validated sentences from the Dutch Common Voice 21.0 dataset \cite{ardila-etal-2020-common} as prompts. Using their respective ground truth utterances as references, 202 clips are generated per speaker. All outputs are resampled from 24 kHz to 16 kHz.

\subsection{ASR models}
To assess the impact of data augmentation for a wider variety of models, we use both a Connectionist Temporal Classification-based (CTC) model and a sequence-to-sequence model. The CTC model is MMS \cite{pratapScalingSpeechTechnology2024}, a wav2vec2.0-based \cite{baevskiWav2vec20Framework2020} multilingual model pretrained on 500,000 hours of speech spanning 1406 languages. We use the \texttt{mms-1b-fl102}\footnote{~\url{https://huggingface.co/facebook/mms-1b-fl102}} checkpoint with the language set to Dutch. Only the adapter is finetuned in our experiments. We conduct experiments both with the Dutch MMS n-gram language model\footnote{~\url{https://huggingface.co/facebook/mms-cclms}} (LM) and with greedy decoding (i.e., without a language model). Not using a language model is expected to hurt performance, because only acoustic information is used. This may for example lead to non-existing words in the transcriptions, a problem which may be further exacerbated by the articulatory imprecision found in oral cancer speech.

In addition, we use Whisper \cite{radfordRobustSpeechRecognition2022}, a sequence-to-sequence model trained on 680,000 hours of speech, supporting 99 languages. Whisper can be seen as an audio-conditional language model, and is known to perform well on Dutch. We use the \texttt{whisper-large-v3}\footnote{~\url{https://huggingface.co/openai/whisper-large-v3}} checkpoint. In our experiments we perform both finetuning of the entire model as well as parameter-efficient finetuning (PEFT) \cite{peft} using low-rank adaptation (LoRA) \cite{huLoRALowRankAdaptation2022}, with the rank set to 32 and $\alpha = 64$.

For both MMS and Whisper, we use a batch size of 32 and finetune for 750 steps with a 100-step warm-up. For MMS, the learning rate is set to $10^{-3}$ and for Whisper it is set to $10^{-5}$. All experiments are conducted on a single 80GB HBM3 NVIDIA H100.

\subsection{Evaluation}
We use the WER as our evaluation metric. All target transcriptions and model outputs are preprocessed by converting to lowercase, removing punctuation, replacing consecutive whitespace characters by a single space, and stripping leading and trailing whitespace. Additionally, numbers and symbols are replaced by their written version using \texttt{num2words}\footnote{~\url{https://github.com/savoirfairelinux/num2words}}, and "é" is replaced by "e". These normalization steps enable a more accurate comparison between Whisper and MMS, as Whisper can output rich text (e.g., punctuation and numbers) while MMS cannot. Replacing numbers by their written variant is necessary, both because the news headlines frequently contain numbers (which MMS cannot output) and to avoid incorrect penalization (e.g., the WER between "2025" and "tweeduizend vijfentwintig" is 200\%, despite being semantically and phonetically identical).

\subsection{Large language model correction}
As an additional experiment, we explore the use of LLMs to correct our models' predictions. We employ GPT-4o, GPT-4o-mini and GPT-3.5-turbo using the OpenAI API, using the following prompt \cite{halpernExplainableReferenceFreeSpeech2025}:
\begin{mdframed}[nobreak=true]
    \textbf{Prompt:} The following is the output of an automatic speech recognition system for an utterance of a speaker with speech pathology in Dutch: [sentence]

    Please correct the sentence. Please put the corrected sentence within square brackets [like this]. If the sentence is already correct, repeat the sentence within square brackets.
\end{mdframed}
This prompt is evaluated for each sentence in the test set. As the OpenAI API's outputs are non-deterministic, we perform the correction three times for each experiment and report the mean of the three runs. For all runs, the temperature is set to 0 and the maximum number of output tokens is set to 400. During our experiments, occasionally a small fraction of requests would result in an error, thus not returning a corrected prediction. For these requests, the original, uncorrected prediction is used to calculate the WER.

\section{Results and discussion}
\begin{table*}[!t]
\centering
\caption{WER results averaged over patient speakers. \textbf{Boldface} indicates best performance per model and dataset. \underline{Underline} indicates best overall performance.}
\label{tab:wer_gpt}
\setlength{\tabcolsep}{3pt}
\resizebox{1.0\textwidth}{!}{
\begin{tabular}{l cccc | cccc | cccc | cccc}
\toprule
 & \multicolumn{4}{c|}{\textbf{MMS}} & \multicolumn{4}{c|}{\textbf{MMS + n-gram}} & \multicolumn{4}{c|}{\textbf{Whisper}} & \multicolumn{4}{c}{\textbf{Whisper LoRA}}\\
\textbf{Dataset} & \textbf{Uncorrected} & \textbf{GPT-4o} & \makecell{\textbf{GPT-4o}\\\textbf{mini}} & \makecell{\textbf{GPT-3.5}\\\textbf{turbo}} & \textbf{Uncorrected} & \textbf{GPT-4o} & \makecell{\textbf{GPT-4o}\\\textbf{mini}} & \makecell{\textbf{GPT-3.5}\\\textbf{turbo}} & \textbf{Uncorrected} & \textbf{GPT-4o} & \makecell{\textbf{GPT-4o}\\\textbf{mini}} & \makecell{\textbf{GPT-3.5}\\\textbf{turbo}} & \textbf{Uncorrected} & \textbf{GPT-4o} & \makecell{\textbf{GPT-4o}\\\textbf{mini}} & \makecell{\textbf{GPT-3.5}\\\textbf{turbo}}  \\
\midrule
No finetuning                               & 44.5 & 35.3 & \textbf{33.9} & 34.7   & 29.1 & \textbf{26.4} & 27.3 & 26.8            & 21.7 & 20.7 & \textbf{20.6} & 20.7                    & 21.7 & 20.7 & \textbf{20.6} & 20.7 \\
All speakers (no augm.)                          & 35.1 & \textbf{19.7} & 19.8 & 20.0   & 19.3 & \textbf{16.7} & 17.3 & \textbf{16.7}   & 20.9 & \textbf{16.7} & 17.1 & 16.9                    & 16.5 & 13.3 & \textbf{13.1} & \textbf{13.1} \\
Patients only (no augm.)                     & 36.1 & \textbf{21.0} & 21.2 & 21.8   & 20.3 & 18.1 & \textbf{17.6} & 17.8            & 19.6 & \textbf{15.5} & \textbf{15.5} & \textbf{15.5}  & 16.3 & \textbf{13.2} & 13.7 & 13.3 \\
\hspace{0.1cm} + Time stretching    & 34.9 & \textbf{19.4} & 20.1 & 20.9   & 19.5 & 17.3 & 17.3 & \textbf{17.1}            & 20.9 & 17.4 & \textbf{17.3} & \textbf{17.3}           & 16.5 & 14.0 & \textbf{13.9} & \textbf{13.9} \\
\hspace{0.1cm} + Speed perturbation   & 36.0 & 19.7 & 19.6 & \textbf{19.4}   & 19.9 & \textbf{17.4} & 17.6 & 17.5            & 19.5 & 15.3 & 15.5 & \textbf{15.2}                    & 16.3 & 13.9 & \textbf{13.3} & 13.4 \\
\hspace{0.1cm} + VTLP                       & 36.2 & 18.8 & 19.1 & \textbf{18.5}   & 19.9 & \textbf{18.2} & \textbf{18.2} & 18.3   & 19.8 & 15.3 & 15.2 & \textbf{15.0}                    & 16.1 & 13.2 & 13.3 & \textbf{\underline{12.9}} \\
\hspace{0.1cm} + kNN-VC                     & 36.4 & \textbf{19.8} & 20.7 & 20.1   & 20.6 & 17.6 & 17.4 & \textbf{17.1}            & 19.7 & 15.3 & \textbf{15.2} & 15.8                    & 16.3 & 13.4 & \textbf{13.3} & 13.4 \\
\hspace{0.1cm} + XTTS                       & 32.7 & 18.2 & 18.2 & \textbf{17.6}   & 17.9 & \textbf{15.1} & 15.4 & 15.3            & 17.3 & 13.8 & 13.4 & \textbf{\underline{12.9}}                    & 16.4 & \textbf{13.3} & 13.6 & 13.4 \\\bottomrule
\end{tabular}
}
\end{table*}
In the following sections, we first discuss the baseline results of the non-finetuned models and those finetuned only on real data. We then compare how using augmented data affects performance. Finally, we discuss how using LLM models for error correction can further influence performance. Results are reported as mean WER across test speakers, and can be found in \autoref{tab:wer_gpt}.

\subsection{Baseline results}
In the baseline experiments, Whisper performs best with a WER of 21.7\%, which is expected, given its relatively strong language model. MMS without a language model presents the poorest performance, as it uses only acoustic information (i.e., no external linguistic knowledge or strong implicit language model). When paired with a simple 5-gram language model (MMS+LM), WER improves by 35\% relative. 

The second baseline consists of models finetuned only on real data from our dataset. We experiment with two approaches: models finetuned on all speakers in the dataset (i.e., including control speech), and models finetuned only on patient speech. This approach was chosen to investigate the effect of including control speech: models may benefit from simply adding more data, or may "forget" typical speech if finetuned only on oral cancer speech. We find that MMS benefits slightly more from including all speakers, but sees large performance increases as a result of either approach, approaching Whisper's performance when paired with a language model. Whisper performs best when finetuned on only patients, and its performance improves especially when finetuned using LoRA (henceforth: Whisper LoRA). 

\subsection{Augmentation results}
For the augmentation experiments, we chose to use only the patient data as our base dataset, due to it achieving the best overall baseline result.

Our time stretching experiments improve performance for MMS, performing on par with the models finetuned on all speakers. For Whisper and Whisper LoRA, using time stretched data sees a regression in performance, also on par with finetuning on all speakers. Using speed perturbation has no significant effect on performance, except for a small decrease in WER for MMS+LM. The results of our time stretching and speed perturbation experiments suggest that the temporal structure of oral cancer speech may not be as distorted as in dysarthric speech. 

VTLP also has no significant influence on performance, except for a small improvement for Whisper LoRA, which, by a small margin, scores the overall best WER of 16.1\%. 

Our voice conversion experiments using kNN-VC result in a small performance deterioration for the MMS models, but have no effect on Whisper. 

For all models except Whisper LoRA, using the TTS-augmented data provides the greatest performance increase, with WER decreasing by $>$10\% relative for all three, possibly because of the added linguistic variety. LoRA has been shown to be effective for finetuning models in a low-resource setting \cite{maoSurveyLoRALarge2024, songLoRAWhisperParameterEfficientExtensible2024}, and therefore Whisper LoRA may not benefit as much from the extra linguistic variety as the other models.

\subsection{LLM correction results}
In our LLM experiments, we investigate the effectiveness of asking an LLM to provide a corrected version of the ASR models' predictions. Across all experiments, we see an improvement in WER. As expected, MMS benefits most, likely due to its lack of a language model causing the uncorrected predictions to contain many non-existing words. 

We observe that while the LLMs reduce the average WER across models by 21.4-26.2\% relative (depending on finetuning data), the non-finetuned models only improve by 10.0\% relative on average. For MMS and MMS+LM this may be due to their poor uncorrected predictions being too difficult to correct. However, a similar lack of improvement is seen for Whisper. Averaged over all LLMs, non-finetuned Whisper's WER sees a 4.7\% relative improvement, while Whisper finetuned on patients sees a much larger 19.1\% relative improvement, despite only a slightly lower uncorrected WER. 

The performance improvement of the LLM correction step is similar for each augmentation method, and XTTS remains the best-performing augmentation method for all models except Whisper LoRA, with a 25.2\% relative improvement on average across all LLMs. The smallest relative performance improvement of 21.7\% is seen for the time stretching experiments. In general, we observe slightly better improvements for lower uncorrected WERs: this is expected, as hypotheses with fewer errors are likely to be easier to correct.

The greatest overall results are again seen for Whisper LoRA, where adding the LLM correction step improves the already similar WERs by around 18\% relative for all augmentation strategies except time stretching. The single best WER of 12.9\% is achieved by GPT-3.5-turbo, for Whisper+XTTS and Whisper LoRA+VTLP. However, for Whisper LoRA, ten of the finetuning strategies score a WER between 12.9\% and 13.3\%, and due to the stochastic nature of the LLM correction, we cannot draw a definitive conclusion as to which approach is best.

Our LLM experiments also enable a comparison between the types of LLM used. We see a slight difference in effect on performance between GPT-4o (22.2\% average relative improvement), GPT-4o-mini (21.9\% average relative improvement) and GPT-3.5-turbo (22.4\% average relative improvement). However, GPT-4o-mini is both the fastest and cheapest model to use.

\section{Limitations of the study}
While our experiments show promising results, the performance metrics must be interpreted with caution. First, our dataset is limited in both total duration and number of speakers. Oral cancer speech is known to be variable, and our findings may therefore not generalize well.

Additionally, due to the computational cost of a leave-one-speaker-out approach and experimenting with multiple models, our study is limited to a single augmentation method per experiment. Moreover, each augmentation method is restricted to a single combination of settings.

Another limitation of the study is the lack of ecological validity of the dataset. ASR has been shown to perform better on read speech than on conversational speech \cite{gablerReconsideringReadSpontaneous2023, sheikhImpactSpeechMode2024, tobinAutomaticSpeechRecognition2024}. Additionally, our dataset is recorded in a sound booth, which is not an accurate depiction of real-world acoustic conditions. Moreover, it means the acoustic conditions are the same for each speaker. ASR models may exploit this. Similar issues have previously been found for other datasets \cite{schuUsingUASpeechTORGO2023, liuCleverHansEffect2024}.

Although we achieve a considerable performance increase by using LLMs for error correction, currently these models are computationally expensive. However, as smaller models improve in performance, local error-correction may become feasible. Lastly, as the models in this work are provided as third-party remote services, they are not a good option when sensitive information is present, such as patient-doctor conversation transcriptions.

\section{Conclusion}
In this paper, we have presented a comparative study of data augmentation methods for Dutch oral cancer speech recognition. We finetuned models on a range of augmented variants of NKI-RUG-UMCG, a Dutch dataset consisting of read oral cancer speech. In our experiments, we found that using text-to-speech augmentation is helpful for most of our ASR models, likely due to the increased linguistic variety it provides in the augmented dataset. The best overall performance is achieved by finetuning Whisper models using LoRA. In addition, we studied how using large language models as an error-correcting post-processing step can further help reduce error rates. We experimented with GPT-4o, GPT-4o-mini and GPT-3.5-turbo, and found that all three LLMs provide a further 20-25\% relative WER decrease for finetuned models, but only a 10\% relative decrease for non-finetuned models. While GPT-3.5-turbo achieves the overall best results, the worst overall LLM (GPT-4o-mini) shows only slightly poorer results.

In future work, we would like to explore combining augmentation methods, such as pairing kNN-VC with a rhythm conversion method. We would also like to explore the use of zero-shot TTS models finetuned for pathological speech, once enough data becomes available to train such models. Furthermore, error-correction using smaller large language models for local use should be investigated to alleviate the privacy concerns of sharing sensitive data with a third party.

\section{Acknowledgements}
The data collected received ethical approval by the Institutional Research Board of the University Medical Center Groningen (NL76137.042.20). All participants provided written informed consent. This work is financed by the Dutch Research Council (NWO) under project number 019.232SG.011.

\bibliographystyle{IEEEtran}
\bibliography{references}

\end{document}